\documentclass[12pt]{article}
\usepackage{a4}

\usepackage{amsfonts}
\usepackage{amsmath}
\usepackage{amssymb}
\usepackage{dsfont}

\newcommand{\bea}{\begin{eqnarray}}
\newcommand{\eea}{\end{eqnarray}}
\newcommand{\be}{\begin{equation}}
\newcommand{\ee}{\end{equation}}
\newcommand{\pa}{\partial}
\newcommand{\nn}{\nonumber}
\newcommand{\la}{\langle}
\newcommand{\ra}{\rangle}
\newcommand{\C}{\odot}

\newcommand{\NS}{{\cal N}_{min}}

\renewcommand{\a}{\alpha}
\renewcommand{\b}{\beta}
\renewcommand{\c}{\gamma}
\renewcommand{\d}{\delta}
\renewcommand{\k}{\kappa}
\newcommand{\e}{\epsilon}
\newcommand{\ve}{\varepsilon}
\newcommand{\s}{\sigma}

\newcommand{\tPhi}{\tilde{\Phi}}

\makeatletter

\@addtoreset{equation}{section}
\makeatother

\def\href#1#2{#2}

\begin{document}

\begin{titlepage}

\begin{center}

\hfill 
\vskip 2.8cm


{\Large \bf Supersymmetric reduced models}\\[3mm]
{\Large \bf with a symmetry based on Filippov algebra}
\vskip 22mm
{\sc Kazuyuki Furuuchi}\footnote{e-mail address: furuuchi@phys.cts.nthu.edu.tw},
\ 
{\sc Dan Tomino}\footnote{e-mail address: tomino@phys.cts.nthu.edu.tw} \ \ \ 
\vskip 10mm
{\sl
National Center for Theoretical Sciences\\
National Tsing-Hua University, Hsinchu 30013, Taiwan, R.O.C.}\\
\noindent{ \smallskip }\\

\vspace{8pt}
\end{center}
\begin{abstract}
Generalizations of the
reduced model of super Yang-Mills theory
obtained by replacing the
Lie algebra structure 
to Filippov $n$-algebra structures are studied.
Conditions for the reduced model actions 
to be supersymmetric are examined.
These models are related with what we call $\NS=2$ super $p$-brane actions.
\end{abstract}

\end{titlepage}

\renewcommand{\thefootnote}{\arabic{footnote}}
\setcounter{footnote}{0}

\section{Introduction}

Gauge symmetry based on Lie algebra 
\cite{Yang:1954ek}
has a
rather long history and 
it has successfully described 
weak and strong interactions in the nature.
The non-Abelian Lie algebra gauge symmetry
on the worldvolume of multiple D-branes
was also a crucial ingredient
in the recent developments in
non-perturbative string theory.
It was also essential in the
matrix model proposals \cite{Banks:1996vh,Ishibashi:1996xs}
which use dimensionally reduced 
super Yang-Mills theory for definition.

Filippov $n$-algebra 
\cite{Filippov}
is a natural generalization
of Lie algebra.
It began to attract wide attention from physicists
recently after it appeared in a candidate model for multiple M2-branes
\cite{Basu:2004ed,Bagger:2006sk,Bagger:2007jr,Gustavsson:2007vu}.

So far studies involving
Filippov 
$n$-algebra in physics have been
largely concentrated
on the Filippov $3$-algebra appearing in the multiple M2-brane model.\footnote{%
With a notable exception of the Nambu bracket \cite{Nambu:1973qe}
which can be used to define a classic example 
of Filippov 3-algebra. 
Quantization of Nambu bracket and/or its application
to brane models have been subjects of interests,
see e.g. \cite{Curtright:2002fd} and references therein.}
It will be interesting to look for other 
situations where 
Filippov 
$n$-algebra plays a role.

In this paper, 
we study generalizations of reduced super Yang-Mills theory
obtained by replacing the Lie algebra structure to 
Filippov 
$n$-algebra, and
examine when the reduced actions are supersymmetric.
Reduced model is a candidate framework for 
a constructive definition of
fundamental theory \cite{Ishibashi:1996xs},
and supersymmetry is expected to be a vital element
in such a framework.

Another motivation for this study comes from 
a trial to relate the
multiple M2-brane action 
with some covariant formalism, possibly 
the single M5-brane action
\cite{Ho:2008nn,Ho:2008ve,Park:2008qe,Furuuchi:2008ki,Bandos:2008fr}
(see also \cite{Bonelli:2008kh}).
In particular, Ref.\cite{Furuuchi:2008ki} studied this issue
from the viewpoint of space-time supersymmetry algebra.
Although results in the above works suggest such a relation,
complete understanding is still missing.
In this paper, we will show that
our reduced models have the same structure
with a covariant Green-Schwarz type supermembrane action 
written in the membrane analogue of 
the Schild action \cite{Schild:1976vq}.
This result will be a useful guide for 
understanding the above issue.

\section{Filippov $n$-algebra}

In this section we briefly review
the necessary ingredients of
Filippov $n$-algebra.
The presentation in this section
closely follows Ref.\cite{Ho:2008bn}.

Filippov $n$-algebra \cite{Filippov},
also known as $n$-Lie algebra,
is a natural generalization of Lie algebra.
(In this paper we will sometimes call it just $n$-algebra for short.)
For a linear space 
${\cal V} = \sum_{a=1}^{\dim {\cal V}} v_a T_a; v_a \in \mathbb{C}$,
Filippov $n$-algebra structure is defined by a multi-linear map
which we call $n$-bracket
$[*,\cdots,*]$ : ${\cal V}^{\otimes n} \rightarrow {\cal V}$
satisfying the following properties:\\
\ \\ 
1. Skew-symmetry:
\be
 \label{skew}
[A_{\s(1)}, \cdots , A_{\s(n)}] = (-1)^{|\s|} [A_1, \cdots, A_n].
\ee
2. Fundamental identity:
\bea
 \label{FI}
&&[A_1,\cdots,A_{n-1},[B_1,\cdots,B_n]] \nn \\
&=&
\sum_{k=1}^n
[B_1,\cdots,B_{k-1},[A_1,\cdots,A_{n-1},B_k],B_{k+1},\cdots,B_n].
\eea
In terms of the basis $T_a$, $n$-algebra
is expressed in terms of the structure constants:
\bea
[T_{a_1},\cdots,T_{a_n}] = i f_{a_1\cdots a_n}{}^b T_b .
\eea

We introduce inner product
as a bi-linear map ${\cal V}\times {\cal V} \rightarrow \mathbb{C}$:
\bea
\la
T_a,T_b
\ra
=h_{ab} .
\eea
The symmetric tensor $h_{ab}$ will be called metric 
of the $n$-algebra in the following.

We impose invariance of the metric
\bea
 \label{invm}
\la [T_{a_1}, \cdots, T_{a_{n-1}},T_b], T_c \ra 
+ \la T_b, [T_{a_1},\cdots,T_{a_{n-1}}, T_c] \ra = 0.
\eea
This implies the tensor
\bea
 \label{antisym}
f_{a_1\cdots a_{n+1}} \equiv f_{a_1\cdots a_n}{}^b h_{b a_{n+1}}
\eea
to be totally anti-symmetric.

We define Hermitian conjugation as follows:
\bea
[A_1,\cdots,A_n]^{\dagger}
=
[A_n^{\dagger},\cdots,A_1^{\dagger}].
\eea

\section{Supersymmetric reduced model actions 
with a symmetry based on Filippov $n$-algebra}

IIB matrix model \cite{Ishibashi:1996xs} is defined as
a large $N$ reduced model of
ten dimensional super Yang-Mills theory. 
Its action is given by
\bea
 \label{IIB}
S =
\frac{1}{4}
\la
[X_I, X_J],
[X^I, X^J]
\ra
+
\frac{1}{2}
\la
\bar{\Psi}, 
\Gamma_{I} [X^{I},\Psi]
\ra .
\eea
Here, $X^I$ $(I = 1,\cdots, 10)$ is a vector in ten dimensional
flat target space-time and
$\Psi$ is a space-time Majorana-Weyl spinor, both
take values in $U(N)$ Lie-algebra.
$\Gamma_I$'s are gamma matrices in ten dimension.
Repeated vector indices are contracted by space-time metric
$\eta_{IJ} = \mbox{diag}(+,-,\cdots,-)$.
We have used the Filippov $n$-algebra
notations ($n=2$ for ordinary Lie algebra)
described in the previous section.
The inner product is given by the invariant trace
of the Lie algebra.

The action (\ref{IIB}) is invariant under the following
supersymmetry transformation:
\bea
 \label{IIBSUSY}
&&\delta X^I = i \bar{\epsilon} \Gamma^I \Psi , \nn \\
&&\delta \Psi = 
 \frac{i}{2} [X^I, X^J] \Gamma_{IJ} \epsilon .
\eea

A natural generalization of the action (\ref{IIB})
based on Filippov $(p+1)$-algebra would be
\bea
 \label{redp}
&&S =
\frac{1}{2(p+1)!}
\la
[X_{I_1}, \cdots, X_{I_{p+1}}]
[X^{I_1}, \cdots, X^{I_{p+1}}]
\ra
\nn \\
&&\qquad \qquad +
\frac{\sigma}{2}
\la
\bar{\Psi}, 
\Gamma_{I_1 \cdots I_{p}} [X^{I_1}, \cdots, X^{I_{p}},\Psi]
\ra .
\eea
Here $\sigma$ 
is a factor $1$ or $i$ 
determined from the Hermiticity of the action.
$X^I$ $(I = 1,\cdots, D)$ 
is a vector in $D$-dimensional
flat target space-time and
$\Psi$ is a space-time spinor, both
take value in $(p+1)$-algebra.
$\Gamma_I$'s are $D$ dimensional gamma matrices
satisfying 
\bea
 \label{gamma}
\Gamma_I \Gamma_J + \Gamma_J \Gamma_I = 2 \eta_{IJ} ,
\eea
where $\eta_{IJ}$ is now $D$-dimensional flat metric
with $\eta_{IJ} = \mbox{diag}(\overbrace{+,\cdots,+}^{t},\overbrace{-,\cdots,-}^{s})$.
We allow the number of the time-like directions to be general $t$.
$\Gamma_{I_1 \cdots I_{p}}$ is an anti-symmetrized product of gamma matrices
with ``strength one".

The action (\ref{redp}) is
invariant under a transformation
\bea
 \label{gauge}
\delta \Phi^a = \Lambda^{a_1\cdots a_p} f_{a_1 \cdots a_p b}{}^a \Phi^b ,
\quad
\Phi^a = X^{I a}, \Psi^a ,
\eea
due to the fundamental identity (\ref{FI}) 
and the invariance of the inner product (\ref{invm}).
This is a natural generalization
of the dimensionally reduced gauge symmetry
of the action (\ref{IIB}).

In this paper, we examine
in which case the following
supersymmetry transformation
\bea
 \label{LSUSY}
&&\delta X^I = c_1 \bar{\epsilon} \Gamma^I \Psi , \nn \\
&&\delta \Psi =
c_2
[X^{I_1}, \cdots, X^{I_{p+1}}] \Gamma_{I_1 \cdots I_{p+1}} \epsilon ,
\eea
leaves the action (\ref{redp}) invariant.
Here, $c_1$ and $c_2$ are coefficients to be adjusted.
We will keep the Filippov algebra to be general,
i.e. we will not use any property specific to 
a particular Filippov algebra.
The conditions we may impose on fermions 
are the standard ones,
i.e. (pseudo-)Majorana condition
and Weyl condition.
We will not consider
projections on fermions 
which break 
the $SO(t,s)$
Lorentz symmetry.
In order for the second term in the action
(\ref{redp}) to be not identically zero,
when we impose Weyl condition on fermions
$t+p$ must be even, and
when fermions are Majorana-spinors
$\Gamma_{I_1 \cdots I_p}C$ must be symmetric
in spinor indices.
Here, $C$ is the charge conjugation matrix.
The properties of gamma matrices and spinors in
diverse dimensions are summarized in appendix \ref{spinors}.

Let us first study the variation of the action 
which has one fermion.
The variation of the second term in the action 
(\ref{redp}) containing one fermion has a form
\bea
 \label{onef}
&&
\la
\bar{\epsilon} 
\Gamma^{J_1 \cdots J_{p+1}}
\Gamma^{I_1 \cdots I_{p}}
[X_{J_1}, \cdots, X_{J_{p+1}}]
[X_{I_1}, \cdots, X_{I_{p}}, \Psi ]
\ra \nn \\
&=&
-
\la
\bar{\epsilon} 
\Gamma^{J_1 \cdots J_{p+1}}
\Gamma^{I_1 \cdots I_{p}}
[X_{I_1}, \cdots, X_{I_{p}},[X_{J_1}, \cdots, X_{J_{p+1}}]]
\Psi
\ra ,
\eea
where we have used the invariance of the 
inner product (\ref{invm}).
One can rearrange the ordering of the 
gamma matrices 
into a sum of 
totally anti-symmetrized gamma matrices
using (\ref{gamma}):
\bea
 \label{asgamma}
\Gamma_{J_1 \cdots J_{p+1}}
\Gamma^{I_1 \cdots I_{p}}
&=&
\Gamma_{J_1 \cdots J_{p+1}}{}^{I_1 \cdots I_{p}} \nonumber \\
&+&
(-)^{p}
\delta^{[I_1}_{[J_1}\Gamma_{J_2 \cdots J_{p+1}]}{}^{I_2 \cdots I_{p}]} \nonumber \\
&+&
\delta^{[I_1}_{[J_1} \delta_{J_2}^{I_2}
\Gamma_{J_2 \cdots J_{p+1}]}{}^{I_3 \cdots I_{p}]} \nonumber \\
&+&
\cdots \nonumber \\
&+&
(-)^{p}
\delta_{[J_1}^{[I_1} \cdots \delta_{J_p}^{I_p]} \Gamma_{J_{p+1}]}  ,
\eea
where the square brackets on Lorentz indices denote 
total anti-symmetrization with appropriate ``strength"
(it will be relevant only for the last term).
On the other hand,
using the 
fundamental identity (\ref{FI})
one can show
\bea
 \label{fromFI}
\Gamma^{I_1\cdots I_r J_1 \cdots J_{r+1}}
[A_1,\cdots,A_{p-r},X_{I_1},\cdots,X_{I_r},
[A_1,\cdots,A_{p-r},X_{J_1},\cdots,X_{J_{r+1}}]] =0 ,
\eea
for $r \ne 0$, with pairs of the same entries $A_1,\cdots, A_{p-r}$.
Due to (\ref{fromFI}) the terms from (\ref{onef}) 
arising from the rearrangement 
of the gamma matrices
(\ref{asgamma})
mostly vanish; only the
$r=0$ term remains which cancels the similar term 
coming from the 
variation of the first term in the action (\ref{redp}).

Next, let us examine 
the variation of the action containing three fermions.
Since
the structure constant $f_{a_1 \cdots a_{p+1}}{}^{b}$
is anti-symmetric in indices 
$a_1 \cdots a_{p+1}$ due to the skew-symmetric property
(\ref{skew}), 
the variation containing three fermions
vanishes when
\bea
 \label{3ferm}
(\Gamma_I)^\a{}_\b (\Gamma^{I J_1 \cdots J_{p-1}})^\c{}_\d
{\Psi}_\b^{[a_1} \bar{\Psi}^{a_2}_\c \Psi_\d^{a_3]}= 0,
\eea
where the square bracket denotes the total anti-symmetrization in the 
$(p+1)$-algebra indices.
(\ref{3ferm}) is equivalent to
\bea
 \label{idC}
(\Gamma_IP)^{\a}{}_{(\b} (\Gamma^{I J_1 \cdots J_{p-1}}P)^{\c}{}_{\d)} = 0 ,
\eea
when $\Psi$'s are complex spinors, where 
$P$ is a chiral projection when $\Psi$'s are Weyl spinors
and $1$ otherwise, and
\bea
 \label{idR}
(\Gamma_ICP)_{\a(\b} (\Gamma^{I J_1 \cdots J_{p-1}}CP)_{\c\d)} = 0 ,
\eea
when $\Psi$'s are (pseudo-)Majorana(-Weyl) spinors.
From an argument similar to the one in \cite{Achucarro:1987nc},
when (\ref{3ferm}) is satisfied it follows that
\bea
 \label{Sredexist}
D-p-1 = \frac{1}{2} n_{f} ,
\eea
where $n_f$ is the spinor size of the fermions $\Psi$ 
counted in the real number.\footnote{%
When Dirac spinors have $n^{\mathbb{C}}_D = \frac{1}{2} n_D^{\mathbb{R}}$
spinor components,
by the size of the spinor counted in real number we mean
$n_D^{\mathbb{R}}$. 
(pseudo-)Majorana condition or Weyl condition
reduces size of spinors by half:
(pseudo-)Majorana spinors and Weyl spinors have size $\frac{1}{2} n_D^{\mathbb{R}}$
and (pseudo-)Majorana-Weyl spinors have size $\frac{1}{4} n_D^{\mathbb{R}}$.}
We provide a proof in the appendix \ref{proofmatch}.
One can also check that (\ref{Sredexist}) is a sufficient condition
for (\ref{3ferm}) to vanish by
expanding the left hand side of (\ref{idC}) or
(\ref{idR}) by complete basis of matrices with indices $\a$ and $\b$. 
We list the cases with $p \geq 2$
when (\ref{3ferm}) is satisfied in Table \ref{tb:Dtp}
($p=1$ case is the ordinary Lie-algebra case 
which can be easily included).
The columns for $\theta$ and $n_\theta$ in the table
are about corresponding super $p$-branes which we will discuss
in section \ref{p-branes}.
\begin{table}[h]
\begin{center}
\begin{tabular}{|c|c|c|c|c|c|c|}
\hline
$D$ & $t$ & $p$ & $\Psi$ & $n_f$ & $\theta$ & $n_{\theta}$ \\
\hline
4 & 2 & 2 & (pseudo-)Majorana-Weyl & 2 &(pseudo-)Majorana & 4 \\
\hline
5 & 2 & 2 & Majorana & 4 & Dirac & 8\\
\hline
5 & 3 & 2 & pseudo-Majorana & 4 & Dirac & 8 \\
\hline
6 & 3 & 3 & (pseudo-)Majorana-Weyl & 4 & Weyl & 8 \\
\hline
\end{tabular}
\caption{The dimension of the target space-time $D$ and 
the number of its time-like dimensions $t$
where the supersymmetric reduced model with $(p+1)$-algebra symmetry
and corresponding Green-Schwarz type super $p$-brane exist.
The column under $\Psi$ is the spinor property of
the fermions in the supersymmetric reduced models
and $n_f$ is the spinor size of $\Psi$,
and the column under $\theta$ is the 
spinor property 
of the space-time spinor fields $\theta$ 
of the corresponding
Green-Schwarz type super $p$-branes and $n_{\theta}$ is
its spinor size.}
\label{tb:Dtp}
\end{center}
\end{table}

We explicitly write down the supersymmetric reduced model action
in the case of $D=4$, $t=2$, $p=2$ with
pseudo-Majorana-Weyl conditions on fermions:
\bea
 \label{Sredp}
S=
\frac{1}{6}
\la
[X_I,X_J,X_K]
[X^I,X^J,X^K]
\ra
+
\frac{1}{2}
\la
\bar{\Psi}\Gamma_{IJ}[X^I,X^J,\Psi]
\ra .
\eea
The supersymmetry transformation is given by
\bea
 \label{LSUSYF}
\delta X^I &=& i \bar{\e} \Gamma^I \Psi, \nn \\
\delta \Psi 
&=& \frac{i}{6} [X^I,X^J,X^K] \Gamma_{IJK} \e.
\eea
The case
with Majorana-Weyl fermions
is similar,
with appropriate modifications 
in the coefficients 
in 
(\ref{LSUSYF}).\footnote{%
Actually in the action (\ref{Sredp}),
the difference between
Majorana-Weyl fermions and
pseudo-Majorana fermions is
just a matter of convention:
The difference arises from the choice
of the charge conjugation matrix
$C_{\eta'=1}$ (for Majorana fermions) 
or $C_{\eta'= -1}$ (for pseudo-Majorana fermions)
in (\ref{Cchoice}) in the appendix \ref{spinors},
which are related as
$C_{\eta'=1} = \Gamma_5 C_{\eta'= -1}$.
Using the Weyl condition on fermions,
one can see that the action for 
pseudo-Majorana-Weyl fermions and that for
Majorana-Weyl fermions are exactly the same.}

\section{Super Poincar\'{e} algebra}

In the previous section we called
the fermionic transformation
(\ref{LSUSYF}) ``supersymmetry transformation", since 
it is an analogue of the supersymmetry of 
the reduced model of super Yang-Mills theory.
However, we haven't shown its relation to 
the standard supersymmetry algebra,
namely super Poincar\'{e} algebra.
Let us examine this point in this section.

We again take the case
$D=4$, $t=2$, $p=2$ 
for explicitly.
Other cases are similar.
In (\ref{Sredp}), the fermions
are pseudo-Majorana-Weyl spinors:
\bea
C\bar{\Psi}^T = \Psi, \quad P_+ \Psi =\Psi ,
\eea
where
\bea
P_\pm \equiv \frac{1 \pm \Gamma_5}{2}, \quad
\Gamma_5 \equiv \Gamma_1 \Gamma_2 \Gamma_3 \Gamma_4  .
\eea
It is important to notice 
that when the 3-algebra
has a central element,
there is a fermionic shift symmetry:\footnote{%
The role of the fermionic shift symmetry in the multiple M2-brane model 
was studied extensively in
\cite{Furuuchi:2008ki}.}
\bea
 \label{NLSUSY}
\delta_+ X^I = 0, \quad \delta_+ \Psi^a = \delta^{a \C} \e_+ ,
\eea
where $\odot$ denotes the central element:
$[T_\C,T_a,T_b] = 0$ for ${}^\forall T_a,T_b$.
The commutation relations of the two fermionic transformations
turn out to be
\bea
 \label{Spp}
(
\delta_+(\e_+^{(1)}) \delta_+(\e_+^{(2)}) -
(1\leftrightarrow 2 )
)
 \Phi &=& 0, \quad \Phi = X^I, \Psi, \\
 \label{trans}
(
\delta_+(\e_+)\delta_-(\e_-)-
\delta_-(\e_-)\delta_+(\e_+)
)
 X^{I a} &=& \delta^{a \C} i \bar{\e}_- \Gamma^I \e_+, \\  
\label{Spm}
(
\delta_+(\e_+)\delta_-(\e_-)-
\delta_-(\e_-)\delta_+(\e_+)
) 
\Psi &=& 0 , \nn \\
 \label{Smm}
(
\delta_-(\e_-^{(1)})\delta_-(\e_-^{(2)}) 
- (1 \leftrightarrow 2)
) 
\Phi_a 
&=& \Lambda_a{}^b \Phi_b ,
\eea
where
\bea
 \label{pMWgt}
\Lambda_a{}^b 
=
i f^{bcd}{}_a X_c^K X_d^L \bar{\e}_-^{(1)} \Gamma_{KL} \e_-^{(2)} .
\eea
Here, $\d_+(\e_+)$ and $\delta_-(\e_-)$ denote the 
fermionic shift (\ref{NLSUSY}) with parameter $\e_+$
and supersymmetry transformation (\ref{LSUSY}) with parameter $\e_-$,
respectively.
(\ref{trans}) is a translation in the target space-time.
Thus the supersymmetry transformation (\ref{LSUSYF})
together with the fermionic shift  (\ref{NLSUSY})
form target space-time super Poincar\'{e} algebra,
modulo the right hand side of (\ref{Smm}) 
which has a form of the
symmetry transformation (\ref{gauge}).
We will use $\NS$ to count the number of supersymmetry
in the unit of the minimal spinor.
In this notation, our model has
$\NS =2$ space-time supersymmetry 
when there is a central element in the algebra,
since
the minimal spinor in $D=4$, $t=2$
is (pseudo-)Majorana-Weyl spinor.
However, as can be seen from (\ref{trans})
it is not possible to construct 
$\NS =1$ space-time super Poincar\'{e} algebra
in four dimension
by using just one minimal spinor. 
In this sense $\NS = 2$ supersymmetry is minimal
in $D=4$, $t=2$ and hence it is what
should be called ${\cal N}=1$ supersymmetry.

So far we have been studying super Poincar\'{e} algebra in
four dimension.
However,
the supersymmetry transformation
(\ref{LSUSYF}) can form super Poincar\'{e} algebra
in three dimension when particular background is chosen.
As an example, let us choose the 3-algebra to be
Nambu-Poisson bracket in ${R^{2,1}}$: 
\bea
[f(y),g(y),h(y)] =
i \e^{ijk} \pa_i f(y) \pa_j g(y) \pa_k h(y)  ,
\eea
\bea
\la f(y) , g(y) \ra
= 
\int d^3y f(y) g(y),
\eea
where $y^i$ $(i=1,2,3)$ are flat coordinates
on ${R^{2,1}}$ and
$\e^{ijk}$ is the Levi-Civita symbol.
We consider following
background configuration:
\bea
 \label{bg}
X^I(y) &=& y^I \quad (I=1,2,3), \nonumber \\
X^4(y) &=& 0  .
\eea
Then, (\ref{pMWgt}) becomes
\bea
 \label{S3}
(
\delta_-(\e_-^{(1)})\delta_-(\e_-^{(2)})
-(1\leftrightarrow 2)
) \tPhi
\sim
\e^{ijk}\bar{\e}_-^{(1)}\Gamma_{jk} \e_-^{(2)} \pa_i \tPhi + \cdots  ,
\eea
where $\tPhi$ are fluctuation of the fields
around the background (\ref{bg}).
To see this is a super Poincar\'{e} algebra in three dimension,
one decomposes gamma matrices and supersymmetry
transformation parameters to those for three dimension.
Then (\ref{S3})
can be rewritten as
\bea
(
\delta_-(\zeta^{(1)}) \delta_-(\zeta^{(2)})
-(1 \leftrightarrow 2)
)\tPhi
\sim
 \bar{\zeta}^{(1)} \gamma^i \zeta^{(2)} \pa_i \tPhi + \cdots ,
\eea
where $\cdots$ can be combined into a form of 
gauge transformation \cite{Ho:2008ve}
and $\gamma^i$ $i=1,2,3$ and $\zeta$ are 
gamma matrices and supersymmetry
transformation parameters in three dimension.
To keep $\tilde{\Phi}=0$ configuration to preserve supersymmetry,
one also needs to combine the fermionic shift (\ref{NLSUSY}) \cite{Ho:2008ve}.
Thus in the background (\ref{bg})
the supersymmetry transformation (\ref{LSUSYF})
appropriately combined with the fermionic shift (\ref{NLSUSY}) 
can be regarded as super Poincar\'{e} symmetry in
three dimension.

\section{Relation to $\NS=2$
super $p$-branes}\label{p-branes}

In this section we show that
our supersymmetric reduced model actions 
can be related to
Green-Schwarz type $\NS=2$ 
super $p$-action 
in the Schild-type form, 
parallel to the relation between
the large $N$
reduced model action of super Yang-Mills theory 
and Green-Schwarz superstring action
\cite{Ishibashi:1996xs}.\footnote{%
Green-Schwarz type supermembrane actions 
with general space-time signatures
have been studied in \cite{Blencowe:1988sk}.
However, there study was restricted to $\NS =1$ case
in our terminology, and
$\NS=2$ supersymmetry which we discuss 
in this paper was not considered there.}
To be explicit, we again take $D=4$, $t=2$, $p=2$ case
as an example.
Discussions are parallel in other cases
listed in Table \ref{tb:Dtp}.

The super $p$-brane action with $p=2$, i.e. 
the supermembrane action is given by
\bea
 \label{mem}
S_{GS}
= 
\int d^3y
\left(
\frac{1}{2} \sqrt{-g}
g^{ij} E_i^I E_j^J \eta_{IJ}
-\frac{1}{2} \sqrt{-g}
+
\epsilon^{ijk}
E_i^A E_j^B E_k^C B_{CBA}
\right),
\eea
where 
we take the worldvolume signature as $(++-)$ and 
$A = (I, \a)$, and
\bea
E_i^I = \pa_i X^I - \frac{i}{2} \bar{\theta}\Gamma^I \pa_i \theta,
\quad
E_i^\a = \pa_i \theta^\a .
\eea
Here, $\theta$ is a pseudo-Majorana spinor in $D=4$, $t=2$ target space-time:
\bea
 \label{Majth}
C\bar{\theta}^T = \theta.
\eea
$B_{ABC}$ is determined from $dB = H$ and $dH = 0$, 
where
\bea
B &=& \frac{1}{3!} E^A E^B E^C B_{ABC} , \nn \\
H &=& \frac{1}{4!}E^A E^B E^C E^D H_{ABCD}, \quad E^A = E_i^A dy^i ,
\eea
and the only non-zero components of 
$H_{ABCD}$ are
those with two spinor and two vector indices:
\bea
H_{\a\b IJ} = - \frac{i}{6} (C^{-1T}\Gamma_{IJ})_{\a\b} .
\eea
The closure of $H$ is equivalent to the identity
\bea
 \label{Gidentity}
(\Gamma_I C)_{(\a\b}\Gamma^{IJ}C_{\c\d)} =0 .
\eea
From this condition
one obtains the matching of on-shell degrees of freedom
between bosons and fermions when 
$2 < p+1 < d$ \cite{Achucarro:1987nc}:
\bea
 \label{bfmatch}
D-(p+1) = \frac{1}{4}  n_{min}\, \NS  ,
\eea
where $n_{min}$ is the dimension of the minimal spinor.
(\ref{bfmatch}) is satisfied 
for $p=2$, $D=4$, $\NS=2$ with pseudo-Majorana-Weyl spinor as 
the minimal spinor; $n_{min}=2$. 
Indeed,
one can show that (\ref{Gidentity}) is satisfied in this case.

The action (\ref{mem}) is invariant 
under the following global space-time supersymmetry transformation:
\bea
 \label{SS}
\delta X^I = \frac{i}{2} \bar{\e} \Gamma^I \theta,
\quad
\delta \theta = \e .
\eea
In terms of the minimal spinor,
the action has $\NS=2$
non-chiral space-time supersymmetry.

The action (\ref{mem}) also has the local
fermionic gauge symmetry:
\bea
 \label{kappa}
\delta X^I = \frac{i}{2} \bar{\theta} \Gamma^I (1+\Gamma) \k ,
\quad
\delta \theta = (1+\Gamma) \k ,
\eea
where
\bea
\Gamma \equiv
\frac{1}{3! \sqrt{-g}} \epsilon^{ijk} E_i^I E_j^J E_k^K \Gamma_{IJK}   .
\eea
The transformation law for the worldvolume metric $g_{ij}$
can be determined as in \cite{Bergshoeff:1987cm}.
To relate the supermembrane action with our reduced model action (\ref{Sredp}),
we fix the fermionic gauge symmetry by the condition\footnote{%
This gauge condition is appropriate for configurations
which break the part of the supersymmetry
generated by $\e_+$.}
\bea
 \label{kgauge}
P_- \theta = 0  ,
\eea
where
\bea
P_{\pm} \equiv \frac{1}{2}(1\pm\Gamma_5), 
\quad \Gamma_5 = \Gamma_1 \Gamma_2 \Gamma_3 \Gamma_4.
\eea
The supersymmetry transformation must be combined
with the global part of the fermionic gauge
transformation to maintain the gauge condition (\ref{kgauge}).
Then the supersymmetry transformation becomes
\bea
\delta X^I = i \bar{\e}_- \Gamma^I \Psi ,
\eea
\bea
\delta \Psi = \e_+ - \Gamma \e_-  ,
\eea
where $\Psi = P_+ \theta$.
After the gauge fixing (\ref{kgauge}), the action takes the form
\bea
 \label{gfGS}
S_{GS} 
= 
\int d^3y
\left(
\frac{1}{2} \sqrt{-g}
g^{ij} \pa_i X^I \pa_j X^J \eta_{IJ}
-\frac{1}{2} \sqrt{-g}
-
\frac{i}{4}
\epsilon^{ijk}
\pa_i X^I \pa_j X^J \bar{\Psi} \Gamma_{IJ} \pa_k \Psi
\right)  .
\eea
This action is classically equivalent to the 
following Schild type action:
\bea
 \label{Schildmem}
S_{Schild}  
&=& 
\frac{1}{2}
\int d^3y \, w(y) \nn \\
&&\left(
- \frac{1}{6}
\{ X^I, X^J,X^K \}\{ X_I, X_J,X_K \} 
- \frac{1}{2}
\bar{\Psi} \Gamma^{IJ} \{X_I, X_J,\Psi \} 
+ 1
\right)  ,
\eea
where $w(y)$ is identified with the volume density and
$\{ *,*,* \}$ is the Nambu-Poisson bracket
\bea
 \label{NPbra}
\{ f,g,h \} \equiv \frac{i}{w(y)}  \epsilon^{ijk}\pa_i f \pa_j g \pa_k h .
\eea
The action (\ref{Schildmem}) can be identified with our 
supersymmetric reduced model (\ref{Sredp}) 
with 3-algebra being the Nambu-Poisson bracket (\ref{NPbra}).
The supersymmetry transformation now becomes
\bea
\delta X^I = i \bar{\e}_- \Gamma^I \Psi ,
\eea
\bea
\delta \Psi = \e_+ + \frac{i}{6}\{X^I,X^J,X^K\}\Gamma_{IJK} \e_-  .
\eea
The fermionic shift symmetry
with the parameter $\e_+$
is identical to (\ref{NLSUSY}),
and the
supersymmetry transformation 
parametrized by $\e_-$ is identical to (\ref{LSUSY}).

Similar discussions go through in
other cases listed in the Table \ref{tb:Dtp}.
$D=5$ models are related with $D=6$ model by
a (formal) double dimensional reduction.
In all cases
$n_\theta$ is the minimal size of the spinor
needed to have super Poincar\'{e} algebra and
it is twice as big as the 
size of the minimal spinor $n_{min}$
in that space-time dimension and signature.
Thus all super $p$-brane actions 
in the Table \ref{tb:Dtp}
have ${\cal N} = 1$, $\NS = 2$ target space supersymmetry.
Note that
the condition for the existence 
of the Green-Schwarz type super $p$-branes
(\ref{bfmatch})
coincides with 
the condition for the existence of
the supersymmetric reduced models
(\ref{Sredexist})
with
$n_f = n_{min}$ and $\NS = 2$, as it should be.

\section{Summary and future directions}

In this paper, we constructed 
supersymmetric reduced model actions with a symmetry 
based on Filippov algebra. 
These models are natural generalizations of the reduced model
of super Yang-Mills theory.
The supersymmetry transformation itself
involves the Filippov algebra structure,
and our models compactly
exhibit interrelation between
supersymmetry and the Filippov algebra symmetry.

The supersymmetric reduced models were related with 
what we call
$\NS=2$ super $p$-brane actions.
In rewriting the super $p$-brane actions
in the form of our reduced models,
there was no truncation of the terms
of the super $p$-brane actions.
Since our models capture the aspects of the symmetries 
in a compact form,
they will provide a good guidance for the
issue of relating the
multiple M2-brane model 
with some covariant formalism
\cite{Ho:2008nn,Ho:2008ve,Park:2008qe,Furuuchi:2008ki,Bandos:2008fr}.
Our models have a nice feature that
the $D$-dimensional Lorentz covariance is manifest.
This was due to $\NS = 2$ supersymmetry
which allowed us to fix the fermionic gauge symmetry
in the Lorentz covariant form (\ref{kgauge}).\footnote{%
While we were completing this work, a paper
\cite{Sato:2009mf} appeared which proposed 
a reduced model with Filippov 3-algebra structure
as a covariant formulation
of M-theory. 
Only the bosonic part was constructed in that paper.
Since M-theory has ${\cal N}={\NS}=1$ supersymmetry
of eleven dimensional space-time 
as opposed to our ${\NS}=2$ models,
one cannot follow our approach in their model.
To keep the covariance
under eleven dimensional Lorentz transformation
in their model
when including terms for supersymmetric completion,
one would need to modify the structure of the model considerably.}
This 
is in contrast to the 
multiple M2-brane model or
super $p$-branes in the light-cone gauge
which have similar algebraic structures \cite{deWit:1988ig,Bergshoeff:1988hw},
and may become an advantage for understanding 
the structure of the space-time
at more fundamental level.
In particular, 
it will be useful for 
describing space-time uncertainty principle covariantly 
\cite{Yoneya:1997gs}.

One of the advantage of reduced models
is that the path integral 
reduces to ordinary integral and
sometimes
explicit integration is possible,
e.g. \cite{Suyama:1997ig,Moore:1998et,Tomino:2003hb}.
Together with the highly symmetric nature of our models,
we may be able to perform the path integral explicitly
and learn quantum aspects of the models with those symmetries.

\section*{Acknowledgments}

We would like to thank
Pei-Ming Ho, 
Yutaka Matsuo,
Jeong-Hyuck Park,
Sheng-Yu Darren Shih 
and Tomohisa Takimi
for useful discussions.
This work is supported in part
by 
National Science Council of Taiwan
under grant No. NSC 97-2119-M-002-001.

\appendix

\section{Gamma matrices and spinors in diverse dimensions}\label{spinors}

In this appendix we summarize 
the properties of gamma matrices and spinors
in diverse dimensions.
See \cite{Kugo:1982bn} for more detail. 

$D$ dimensional gamma matrices $\Gamma_I$ ($I = 1 , \cdots, D$) 
satisfy
\bea
 \label{appgamma}
\Gamma_I \Gamma_J + \Gamma_J \Gamma_I = 2 \eta_{IJ} ,
\eea
where $\eta_{IJ}$ is $D$-dimensional flat metric
with $\eta_{IJ} = \mbox{diag}(\overbrace{+,\cdots,+}^{t},\overbrace{-,\cdots,-}^s)$.
\bea
\Gamma_I^{\dagger} = 
\left\{
\begin{array}{l} 
\, \, \, \, \Gamma_I \quad (I = 1,\cdots, t)  \nonumber \\
 - \Gamma_I \quad (I = t+1,\cdots, D) .
\end{array}
\right.  
\eea
The charge conjugation matrix $C$ is characterized by the property
\bea
 \label{Cchoice}
C^{-1} \Gamma_I C = \eta' \Gamma_I^T
\quad (\eta' = \pm 1)  .
\eea
For even $D$ either sign of $\eta'$ can be chosen,
but for odd $D$ it is fixed to $\eta' = \frac{D(D-1)}{2}$.
\bea
C^T = \ve' C, \quad C^\dagger C = 1  ,
\eea
it follows that
\bea
\Gamma_{I_1\cdots I_r}C =
(\eta')^r \ve' (-)^{\frac{r(r-1)}{2}}
(\Gamma_{I_1\cdots I_r}C)^T  ,
\eea
where $\Gamma_{I_1\cdots I_r}$ is totally anti-symmetrized
product of gamma matrices with ``strength one".
The rank $r$ of symmetric and anti-symmetric 
$\Gamma_{I_1 \cdots I_r}C$ for the relevant cases
are listed in Table \ref{tb:SymG}.

To define Dirac conjugation, we introduce
\bea
\Gamma_0 \equiv \Gamma_1 \cdots \Gamma_t  .
\eea
It satisfies
\bea
\Gamma_0 \Gamma_0^\dagger = 1, \quad \Gamma_0^\dagger = (-)^{\frac{t(t-1)}{2}} \Gamma_0,
\eea
\bea
\Gamma_0 \Gamma_I^\dagger \Gamma_0^\dagger = (-)^{t+1} \Gamma_I  .
\eea
Dirac conjugate field $\bar{\psi}$ of $\psi$ is defined as
\bea
\bar{\psi} = \psi^\dagger \Gamma_0^{-1}  .
\eea
Let us introduce following matrix $B$:
\bea
&&B^{-1}\Gamma_I B = \eta \Gamma^{\ast}_I \quad (\eta = \pm 1) , \nonumber \\
&&B^T = \ve B, \quad  B^\dagger B = 1  .
\eea
Charge conjugate field $\psi^c$ of $\psi$ is defined by
\bea
\psi^c = C\bar{\psi}^T = B \psi^{\ast}  .
\eea
It follows that
\bea
B = C\Gamma_0^{\ast}  ,
\eea
and
\bea
\eta = \eta'(-)^{t+1}, \quad
\ve = \ve' (\eta')^{t}(-)^{\frac{t(t-1)}{2}} .
\eea
(pseudo-)Majorana spinors $\psi_M$ satisfy
\bea
 \label{Maj}
\psi_M = C\bar{\psi}_M^T = B\psi_M^*.
\eea
This is possible only when $\varepsilon = +1$
since (\ref{Maj}) implies
$BB^* =1$.

When $D$ is even, we can define $\Gamma_{D+1}$ as 
\bea
\Gamma_{D+1} = i^{(s-t)/2} \Gamma_1 \Gamma_2 \cdots \Gamma_{D},
\eea
which satisfies
\bea
\Gamma_{D+1}^2 = 1, \quad \Gamma_{D+1}^\dagger = \Gamma_{D+1}.
\eea
$\Gamma_{D+1}$ has following $B$-conjugation property:
\bea
 \label{Bconj}
B^{-1} \Gamma_{D+1} B = (-)^{(s-t)/2} \Gamma_{D+1}^*
\eea
Weyl spinors $\psi_\pm$ satisfy
\bea
\Gamma_{D+1} \psi_\pm = \pm \psi .
\eea
However this is compatible with the (pseudo-)Majorana condition
(\ref{Maj}) only if
\bea
(-)^{(s-t)/2} = 1,
\eea
i.e. $s-t=0$ mod $4$.
The values of $s-t$
when (pseudo-)Majorana(-Weyl) spinors exist are listed
in Table \ref{tb:pMW}.

\begin{table}[h]
\begin{center}
\begin{tabular}{|c|c|c|c|c|}
\hline
$D$ & $\eta'$ & $\varepsilon'$ & $r$ of symmetric $\Gamma_{I_1 \cdots I_r}C$ & $r$ of anti-symmetric $\Gamma_{I_1\cdots I_r}C$ \\
\hline
4 & $+$ & $-$  & 2,3 & 0,1,4 \\
\cline{2-5}
  & $-$ & $-$ & 1,2 & 0,3,4 \\
\hline
5 & $+$ & $-$ & 2,3 & 0,1,4 \\
\hline
6 & $+$ & $-$  & 2,3,6 & 0,1,4,5 \\
\cline{2-5}
  & $-$ & $+$ & 0,3,4 & 1,2,5,6 \\
\hline
7 & $-$ & $+$ & 0,3,4 & 1,2,5,6 \\
\hline
8 & $+$ & $+$ & 0,1,4,5,8 & 2,3,6,7 \\
\cline{2-5}
  & $-$ & $+$ & 0,3,4,7,8 & 1,2,5,6 \\
\hline  
\end{tabular}
\caption{The rank $r$ of symmetric and anti-symmetric
$\Gamma_{I_1 \cdots I_r}C$}
\label{tb:SymG}
\end{center}
\end{table}

\begin{table}[h]
\begin{center}
\begin{tabular}{|c|c|c|c|}
\hline
 & $\eta$ & $\varepsilon$ & $s-t$ mod $8$ \\
\hline
Majorana & $+$ & $-$ & $1,2,8$ \\
\hline
pseudo-Majorana & $+$ & $+$ & $6,7,8$ \\
\hline
Majorana-Weyl & $-$ & $+$ & $8$ \\
\hline
pseudo-Majorana-Weyl & $+$ & $+$ & $8$ \\
\hline
\end{tabular}
\caption{The values of $s-t$ where Majorana, pseudo-Majorana, Majorana-Weyl and pseudo-Majorana-Weyl spinors exist.}
\label{tb:pMW}
\end{center}
\end{table}

\section{Proof of Eq.(\ref{Sredexist})}\label{proofmatch}

In this appendix
we show Eq.(\ref{Sredexist})
\bea
 \label{appSredexist}
D-p-1 = \frac{1}{2} n_{f} 
\eea
follows from Eq.(\ref{3ferm}):
\bea
 \label{app3ferm}
(\Gamma_I)^\a{}_\b (\Gamma^{I J_1 \cdots J_{p-1}})^\c{}_\d
{\Psi}_\b^{[a_1} \bar{\Psi}^{a_2}_\c \Psi_\d^{a_3]}= 0.
\eea

As in \cite{Achucarro:1987nc},
when the fermion $\Psi^a$ is a complex spinor
we define a spinor $\Upsilon^a$ by
\bea
\Upsilon^a = 
\left(
\begin{array}{c}
P \Psi^a \\
\overline{P \Psi^a}^T
\end{array}
\right).
\eea
$P$ is a chirality projection
if $\Psi$ is a Weyl spinor and 
the identity matrix otherwise.
We define symmetric matrices
$\Sigma_I$, $\tilde{\Sigma}_I$ by
\bea
\Sigma_I =
\left(
\begin{array}{cc}
0 & \Gamma_I^T \\
\Gamma_I & 0
\end{array}
\right),
\quad 
\tilde{\Sigma}_I =
\left(
\begin{array}{cc}
0 & \Gamma_I  \\
\Gamma_I^T & 0
\end{array}
\right),
\eea
which satisfy
\bea
\tilde{\Sigma}_I \Sigma_J +
\tilde{\Sigma}_J \Sigma_I 
= 2 \eta_{IJ}.
\eea
We define
\bea
Z =
\left(
\begin{array}{cc}
\mathds{1} & 0 \\
0 & \mathds{1} 
\end{array}
\right), \quad
\left(
\begin{array}{cc}
0 & - \mathds{1}\\
\mathds{1} & 0
\end{array}
\right), \quad
\left(
\begin{array}{cc}
-\mathds{1} & 0 \\
0 & \mathds{1} 
\end{array}
\right), \quad
\left(
\begin{array}{cc}
0 & \mathds{1}\\
\mathds{1} & 0
\end{array}
\right), 
\eea
for $p=1$ mod $4$,
$p=2$ mod $4$,
$p=3$ mod $4$,
$p=4$ mod $4$, respectively.

When $\Psi^a$ is (pseudo-)Majorana(-Weyl)
we define
\bea
\Sigma_I = \Gamma_I C , \quad
\tilde{\Sigma}_I =  C^{-1} \Gamma_I, 
\quad
\Upsilon^a = P \Psi^a, 
\quad Z = \mathds{1}.
\eea
Therefore Eq.(\ref{app3ferm}) is equivalent to
\bea
 \label{3fermB}
\Sigma_I \Upsilon^{[ a_1}
\Upsilon^{a_2 T} Z \Sigma^{IJ_1 \cdots J_{p-1}} \Upsilon^{a_3 ]} = 0 ,
\eea
where the square bracket denotes total anti-symmetrization
in $(p+1)$-algebra indices
and $\Sigma^{IJ_1 \cdots J_{p-1}}$ is defined as
\bea
\Sigma^{IJ_1 \cdots J_{p-1}} &=& \Sigma^{[I} \tilde{\Sigma}^{J_1} \cdots \tilde{\Sigma}^{J_{p-2}}\Sigma^{J_{p-1}]}
\quad \mbox{for odd $p$}, \nn \\
\Sigma^{IJ_1 \cdots J_{p-1}} &=& \tilde{\Sigma}^{[I} \Sigma^{J_1} \cdots \tilde{\Sigma}^{J_{p-2}}\Sigma^{J_{p-1}]}
\quad \mbox{for even $p$},
\eea
where the square bracket denotes total anti-symmetrization
in the Lorentz indices.
Since we have doubled the size of the
spinors when the fermions $\Psi$ are complex spinors, 
we can always go to a real basis
by a similarity transformation.
Therefore (\ref{3fermB})
is equivalent to
\bea
 \label{Srede2}
(\Sigma_I \mathds{P})_{\a(\b} (Z \Sigma^{IJ_1 \cdots J_{p-1}}\mathds{P})_{\c\d)} =0,
\eea
where for complex spinors
\bea
\mathds{P} = 
\left(
\begin{array}{cc}
P & 0 \\
0 & \tilde{P}^T
\end{array}
\right)
\eea 
with
$\tilde{P}=P$ for $t$ even and
$\tilde{P}=1-P$ for $t$ odd for Weyl spinors
and $P = \tilde{P}=1$ otherwise,
and $\mathds{P} = P$ for 
(pseudo-)Majorana spinors.
Contracting (\ref{Srede2}) with
$(\tilde{\Sigma}^K)^{\b\a}$
we obtain
\bea
 \label{cntrct}
n_f (Z \Sigma^{KJ_1 \cdots J_{p-1}}\mathds{P})_{\c\d}
+ 
2 (Z \Sigma_{IJ_1 \cdots J_{p-1}} \tilde{\Sigma}^K \Sigma^I \mathds{P})_{\c\d} = 0 ,
\eea
where $n_f$ is the spinor size of fermions $\Psi$ counted in real number.
From (\ref{cntrct}) we obtain
\bea
(n_f - 2(D-p-1)) (Z \Sigma^{K J_1 \cdots J_{p-1}}\mathds{P})_{\c \d} = 0.
\eea
Thus we have obtained Eq.(\ref{appSredexist}):
\bea
 \label{Sredeapp}
D-p-1  = \frac{1}{2} n_f 
\eea
as a necessary condition for Eq.(\ref{app3ferm}) to vanish.
One can check that it is also a sufficient condition.

The derivation of Eq.(\ref{bfmatch}) is similar,
the main difference is 
the spinor size $n_{\theta}$ 
of the space-time spinor field $\theta$
and
the fact that
in (\ref{Srede2}) only three spinor indices are
symmetrized whereas
in the case of super $p$-brane 
the closure of $H$ leads to a condition
\bea
 \label{Sp}
(\Sigma_I \mathds{P})_{(\a\b} (Z \Sigma^{IJ_1 \cdots J_{p-1}}\mathds{P})_{\c\d)} = 0 ,
\eea
i.e. four spinor indices are symmetrized.
From (\ref{Sp})
one obtains \cite{Achucarro:1987nc}
\bea
 \label{matchapp}
D-p-1  = \frac{1}{4} n_{\theta} ,
\eea
for $2 < p+1 < D$.
In all the cases listed in the Table \ref{tb:Dtp},
$n_f = n_{min}$ and $n_\theta = n_{min} \times \NS$ with $\NS =2$.
The difference of the factors
$\frac{1}{2}$ and $\frac{1}{4}$ in (\ref{Srede2}) and (\ref{matchapp})
is 
a consequence
of the fact that in (\ref{Srede2})
three spinor indices were symmetrized
whereas in (\ref{Sp}) four spinor indices were symmetrized.
Since the supersymmetric reduced model actions
are obtained after fixing the fermionic gauge symmetry
of the super $p$-brane actions
which reduces the degrees of freedom of $\theta$
by half, i.e. $n_f = \frac{1}{2} n_\theta$,
this difference of the factors
is what it should be.

\bibliography{flppvrefs}
\bibliographystyle{utphys}

\end{document}